\documentclass[prl,aps,showpacs,nofootinbib,superscriptaddress,tightenlines,twocolumn]{revtex4}

\usepackage{epsfig}
\usepackage[latin1]{inputenc}
\usepackage{float,amsmath}
\usepackage{graphicx}

\newcommand{\beq}{\begin{equation}}
\newcommand{\eeq}{\end{equation}}
\newcommand{\bqa}{\begin{eqnarray}}
\newcommand{\eqa}{\end{eqnarray}}

\def\lsim{\mathrel{\rlap{\lower4pt\hbox{$\sim$}}
    \raise1pt\hbox{$<$}}}                
\def\gsim{\mathrel{\rlap{\lower4pt\hbox{$\sim$}}
    \raise1pt\hbox{$>$}}}                

\begin{document}

\title{Gluon Thermodynamics at Intermediate Coupling}

\author{Jens O. Andersen}
\affiliation{Department of Physics, Norwegian University of Science
and Technology, Trondheim, Norway}
\author{Michael Strickland}
\affiliation{Department of Physics, Gettysburg College, Gettysburg, Pennsylvania
17325, USA;\\
Frankfurt Institute for Advanced Studies,  D-60438 Frankfurt am Main, Germany}
\author{Nan Su}
\affiliation{Frankfurt Institute for Advanced Studies,  D-60438 Frankfurt 
am Main, Germany}

\begin{abstract}
We calculate the thermodynamic functions of Yang-Mills theory to three-loop order using
the hard-thermal-loop perturbation theory reorganization of finite temperature
quantum field theory.  We show that at three-loop order hard-thermal-loop perturbation theory
is compatible with lattice results for the pressure, energy density, and entropy down to 
temperatures $T \sim 2-3\;T_c$.  
\end{abstract}

\maketitle \newpage


The goal of ultrarelativistic heavy-ion collision experiments is to generate energy
densities and temperatures high enough to create a plasma of quarks
and gluons called the quark-gluon plasma.  One of the chief theoretical
questions which has emerged in this area is whether it is more appropriate to
describe the state of matter generated during these collisions using weakly-coupled
quantum field theory or a strong-coupling formalism based on AdS/CFT correspondence.  
Early data from the Relativistic Heavy Ion Collider (RHIC) at Brookhaven National Labs 
indicated that the state of matter created there behaved more like a 
fluid than a plasma and that this ``quark-gluon fluid'' was
strongly coupled \cite{rhicexperiment}.  

In the intervening years, however, work on the perturbative side has shown that 
observables like jet quenching  \cite{pert} and elliptic flow 
\cite{Xu:2007jv} can also be described using a perturbative formalism.  Since in
phenomenological applications predictions are complicated by the 
modeling required to describe, for example, initial state effects, the space-time evolution
of the plasma, and hadronization of the plasma, there are significant theoretical uncertainties
remaining.  
Therefore, one is hard put to conclude whether the plasma is strongly or weakly coupled 
based solely on RHIC data.  To have a cleaner testing ground one can compare theoretical 
calculations with results from lattice quantum chromodynamics (QCD).

Looking forward to the upcoming heavy-ion experiments scheduled to take place at the 
Large Hadron Collider (LHC) at the European Laboratory for Particle Physics (CERN) it is 
important to know if, at the higher temperatures generated, one expects a strongly-coupled (liquid) or 
weakly-coupled (plasma) description to be more appropriate.  At RHIC, initial temperatures 
on the order of one to two times the QCD critical temperature,
$T_c \sim 190$ MeV, were generated.  At LHC, initial temperatures on
the order of $4-5\;T_c$ are expected.  The key question is, will the generated matter behave more
like a plasma of quasiparticles at these higher temperatures?

In this Letter we discuss the calculation of thermodynamic functions of a gas of gluons at 
phenomenologically relevant temperatures.  We present results at leading order (LO),
next-to-leading order (NLO), and next-to-next-to-leading order (NNLO) and compare with
available lattice data \cite{Boyd:1996bx} for the thermodynamic functions of
SU(3) Yang-Mills theory.  The calculation is based on a reorganization of the theory around
hard-thermal-loop (HTL) quasiparticles.  Our results 
indicate that the lattice data at temperatures $T \gsim 2 - 3\;T_c$ 
are consistent with the quasiparticle picture.  This is a non-trivial result
since, in this temperature regime, the QCD 
coupling constant is neither infinitesimally weak nor infinitely strong with 
$g_s \sim 2$, or equivalently $\alpha_s = g_s^2/(4\pi) \sim 0.3$.  Therefore, we 
have a crucial test of the quasiparticle picture in the intermediate coupling regime.

The calculation of thermodynamic functions
using weakly-coupled quantum field theory has a long history \cite{pertthermo}.  The QCD 
free energy is known up to order $g_s^6 \log(g_s)$; 
however, the resulting weak-coupling
approximations do not converge at phenomenologically relevant couplings. 
For example, simply comparing
the magnitude of low-order contributions to the QCD
free energy with three quark flavors ($N_f=3$), 
one finds that the $g_s^3$ contribution is smaller than the $g_s^2$
contribution only for $g_s \lsim 0.9$ ($\alpha_s  \lsim 0.07$) which
corresponds to a temperature of $T \sim 10^5 $ GeV $\sim 5 \times 10^5 \, T_c$. 

The poor convergence of finite-temperature perturbative expansions of 
thermodynamic functions stems from the fact that at high temperature
the classical solution is not described by massless gluonic
states.  Instead one must include plasma effects such as the screening
of electric fields and Landau damping via a self-consistent hard-thermal-loop resummation.
There are several ways of systematically 
reorganizing the perturbative expansion \cite{reorg}. Here we will 
present a new NNLO calculation which uses the 
hard-thermal-loop perturbation theory
(HTLpt) method \cite{htlpt1,htlpt2,qednnlo} and compare with previously
obtained LO and NLO results.

The basic idea of the technique is to add and subtract an effective mass term from 
the bare Lagrangian and to associate the added piece with the 
free part of the Lagrangian and the subtracted piece with the interactions \cite{spt1,spt2}.
However, in gauge theories, one cannot simply add and subtract a local mass
term since this would violate gauge invariance.
Instead, one adds and subtracts an HTL 
improvement term which modifies the propagators and
vertices self-consistently so that the reorganization is 
manifestly gauge invariant \cite{Braaten:1991gm}.

\vspace{3mm}
\noindent
\textsc{Formalism}~: 
The Lagrangian density for SU($N_c$) Yang-Mills theory in Minkowski space is
\begin{eqnarray}
{\cal L}_{\rm YM}&=&
-{1\over4}F_{\mu\nu}F^{\mu\nu} 
+{\cal L}_{\rm gf}
+{\cal L}_{\rm gh}
+\Delta{\cal L}_{\rm YM}\;.
\label{L-YM}
\end{eqnarray}
Here the field strength is 
$F^{\mu\nu}=\partial^{\mu}A^{\nu}-\partial^{\nu}A^{\mu} + i g_s [A^\mu,A^\nu]$,
with $A^\mu$ an element of the SU($N_c$) gauge group.
The ghost term ${\cal L}_{\rm gh}$ depends on the gauge-fixing term
${\cal L}_{\rm gf}$.  We use $\overline{\rm MS}$ dimensional regularization with a
renormalization scale $\mu$ and covariant gauge fixing
${\cal L}_{\rm gf}=-\left(\partial_{\mu}A^{\mu}\right)^2/(2\xi)$ where
$\xi$ is the gauge parameter.  HTLpt is gauge-fixing independent, therefore, all 
results shown below are independent of
the gauge-fixing parameter. The independence of the results on the gauge
parameter was explicitly demonstrated in general covariant and Coulomb gauges in Ref.~\cite{htlpt2}.

HTLpt is a reorganization
of the perturbation
series for thermal gauge theories. In the case of Yang-Mills theory, 
the Lagrangian density is written as
\begin{eqnarray}
{\cal L}= \left({\cal L}_{\rm YM}
+ {\cal L}_{\rm HTL} \right) \Big|_{g_s \to \sqrt{\delta} g_s}
+ \Delta{\cal L}_{\rm HTL}\,.
\label{L-HTLQCD}
\end{eqnarray}
where $\Delta{\cal L}_{\rm HTL}$ collects counterterms necessary
to account for additional divergences introduced by ${\cal L}_{\rm HTL}$.
The HTL improvement term is
\begin{eqnarray}
{\cal L}_{\rm HTL}=-{1\over2}(1-\delta) \, m_D^2 \, {\rm Tr}\left(
F_{\mu\alpha}\left\langle {y^{\alpha}y^{\beta}\over(y\cdot D)^2}
	\right\rangle_{\!\!y}F^{\mu}_{\;\;\beta}
\right) \, ,
\label{L-HTL}
\end{eqnarray}
where $D^{\mu}=\partial^{\mu}+ig_sA^{\mu}$ is the covariant derivative,
$y^{\mu}=(1,\hat{{\bf y}})$ is a light-like four-vector,
and $\langle\ldots\rangle_{ y}$
represents an average over the directions
of $\hat{{\bf y}}$.  
The term~(\ref{L-HTL}) has the form of the effective Lagrangian
that would be induced by
a rotationally-invariant ensemble of color-charged sources in the
Eikonal approximation. The free parameter $m_D$ can be identified with the
Debye screening mass, but is not assumed to be $m_D \sim g_s T$ at leading order.
HTLpt is defined by treating
$\delta$ as a formal expansion parameter and expanding in a power series
in $\delta$ around $\delta=0$.  This generates loops with fully dressed propagators and
vertices and also automatically generates
the counterterms necessary to remove the dressing as one proceeds to 
higher loop orders \cite{htlpt1,htlpt2,qednnlo}.  

\begin{figure}[t]
\begin{center}
\includegraphics[width=7.9cm]{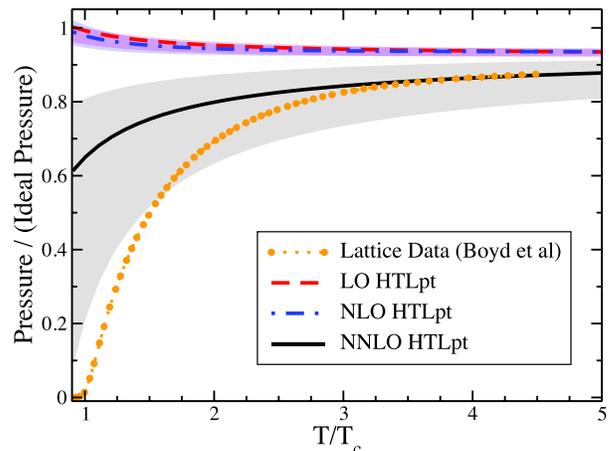}
\end{center}
\vspace{-8mm}
\caption{Comparison of LO, NLO, and NNLO predictions for the scaled pressure 
with SU(3) pure-glue lattice data from Boyd et al. \cite{Boyd:1996bx}.  Shaded
bands show the result of varying the renormalization scale $\mu$ by a factor of two
around $\mu = 2 \pi T$.}
\label{fig:pressure}
\end{figure}

\begin{figure}[t]
\begin{center}
\includegraphics[width=7.9cm]{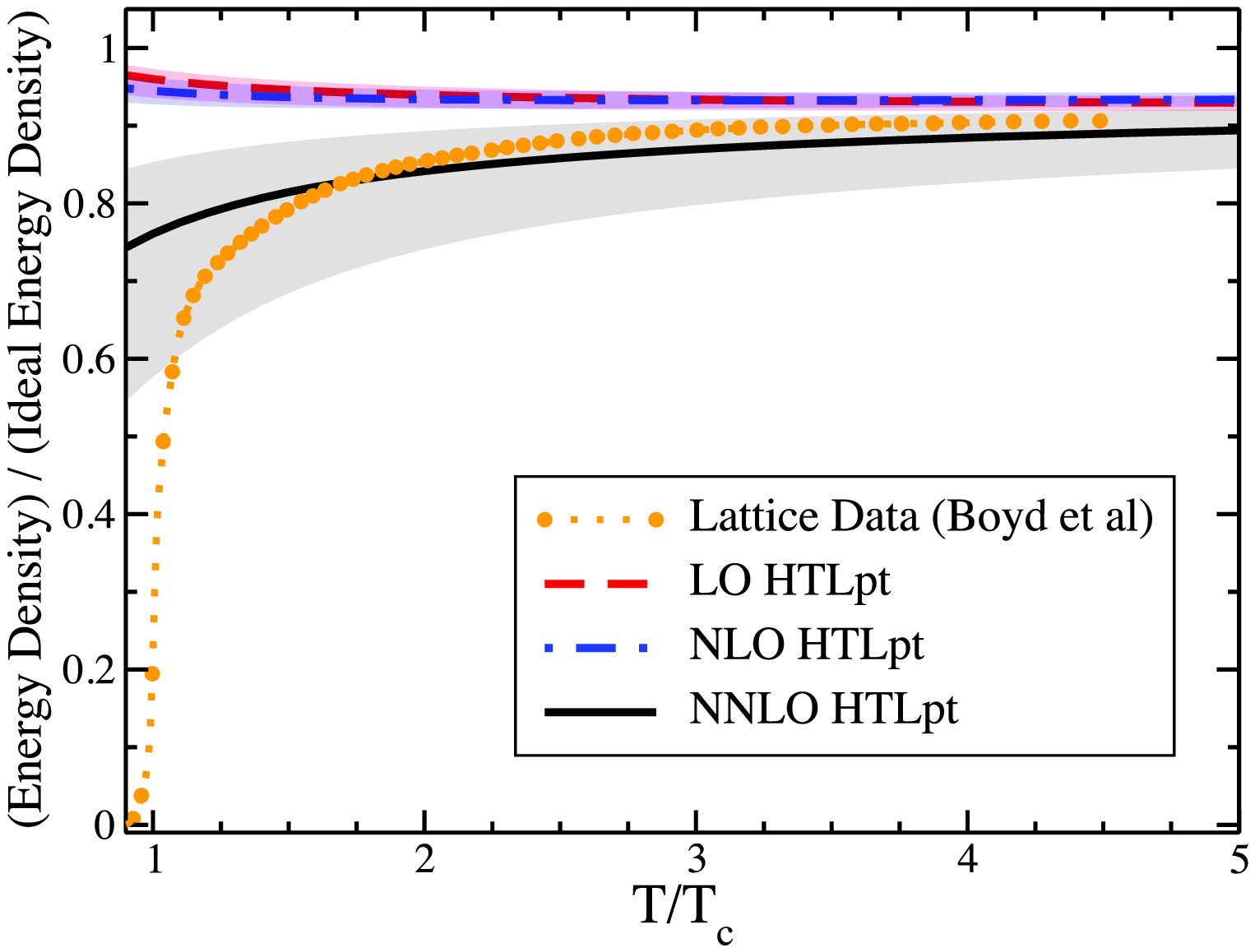}
\end{center}
\vspace{-8mm}
\caption{Comparison of LO, NLO, and NNLO predictions for the scaled energy density 
with SU(3) pure-glue lattice data from Boyd et al. \cite{Boyd:1996bx}.  Shaded
bands show the result of varying the renormalization scale $\mu$ by a factor of two
around $\mu = 2 \pi T$.}
\label{fig:energy}
\end{figure}

\begin{figure}[t]
\begin{center}
\includegraphics[width=7.9cm]{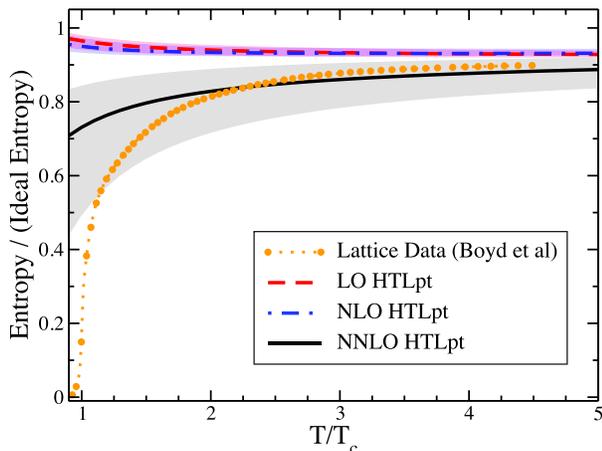}
\end{center}
\vspace{-8mm}
\caption{Comparison of LO, NLO, and NNLO predictions for the scaled entropy 
with SU(3) pure-glue lattice data from Boyd et al. \cite{Boyd:1996bx}.  Shaded
bands show the result of varying the renormalization scale $\mu$ by a factor of two
around $\mu = 2 \pi T$.}
\label{fig:entropy}
\end{figure}

If the expansion in $\delta$ could be calculated to all orders,
the final result would not depend on $m_D$.
However, any truncation of the expansion in $\delta$ produces results
that depend on $m_D$.  We will first obtain the thermodynamic potential 
$\Omega(T,\alpha_s,m_D,\mu,\delta=1)$ which is a function of the
mass parameter $m_D$.  A prescription is then required to determine $m_D$
as a function of $T$ and $\alpha_s$.  The canonical way to fix the Debye mass
in HTLpt is to require the thermodynamic potential to satisfy a variational
equation; however, at NNLO this results in a complex-valued 
Debye mass which causes the thermodynamic functions to also become
complex.  The same problem occurs in QED \cite{qednnlo} and scalar theories \cite{spt1}
and in the case of gauge theories is 
most likely due to the expansion we perform of the resulting integrals in
$m_D/T$ (see next section).  Due to
the complexity of the variational Debye mass, here we will use a NLO perturbative
mass prescription detailed below.  This prescription guarantees that the Debye
mass and hence thermodynamic functions are real-valued at all temperatures.

After having fixed the Debye mass as a function of $T$ and $\alpha_s$, 
the free energy, ${\cal F}$,
is obtained by evaluating the thermodynamic potential at the appropriate
value of the Debye mass.  The pressure, energy density, and entropy are
then evaluated using standard thermodynamic relations 
$
{\cal P} = -{\cal F}, 
{\cal E} = {\cal F} - T \frac{d {\cal F}}{d T}, 
{\cal S} = - \frac{d {\cal F}}{d T}.
$

\vspace{3mm}
\noindent
\textsc{Thermodynamic Potentials}~:
In this section we present the final renormalized 
thermodynamic potential at orders
$\delta^0$ (LO), $\delta^1$ (NLO), and $\delta^2$ (NNLO).  
The LO and NLO results were
first obtained in \cite{htlpt2} and we list them here for 
completeness.  The thermodynamic potentials are 
computed using a dual expansion in $g_s$ and $m_D$
which assumes that at leading order $m_D/T$ is ${\cal O}(g_s)$.
We then only include terms which contribute naively
through order $g_s^5$ \cite{htlpt2}.  This dual truncation
is not necessary in principle; however, in practice it makes
the calculation tractable.  We note that the expansion in
$m_D/T$ does not spoil the gauge invariance of our final
results since the gauge parameter dependence cancels prior
to the expansion in $m_D/T$.

We do not list the renormalization 
counterterms necessary but mention that, as in the case 
of NNLO HTLpt QED \cite{qednnlo}, only systematic vacuum, mass, and coupling 
constant counterterms are 
necessary to renormalize the thermodynamic potential.
The thermodynamic potentials listed below are gauge
invariant.  Full details of the Yang-Mills NNLO calculation will be presented
elsewhere \cite{forthcoming}; however, we note that the calculation
is similar to the one presented in Ref.~\cite{qednnlo}.

\vspace{2mm}
\noindent
{\em Leading~order}~:
The renormalized LO thermodynamic potential is \cite{htlpt2}
\begin{eqnarray}
\frac{\Omega_{\rm LO}}{{\cal F}_{\rm ideal}} &\!\!=\!\!& 
1 - {15 \over 2} \hat m_D^2
+ 30 \, \hat m_D^3 
\nonumber \\ && \hspace{2mm}
+ {45 \over 4}
\left( \log {\hat \mu \over 2}
	- {7 \over 2} + \gamma + {\pi^2\over 3} \right)
	\hat m_D^4   \;,
\label{Omega-LO}
\end{eqnarray}
where ${\cal F}_{\rm ideal} = -(N_c^2-1)\pi^2T^4/45$
is the free energy of an ideal gas of non-interacting gluons,
$\gamma$ is the Euler-Mascheroni constant, and we have
introduced the dimensionless parameters $\hat m_D = m_D /( 2 \pi T)$
and $\hat\mu = \mu /( 2 \pi T)$.

\vspace{2mm}
\noindent
{\em Next-to-leading~order}~:
The renormalized NLO thermodynamic potential is \cite{htlpt2}
\begin{eqnarray}
\frac{\Omega_{\rm NLO}}{{\cal F}_{\rm ideal}} 
&=& 1 - 15 \hat m_D^3
- {45 \over 4} \left( \log{\hat \mu \over 2}
		- {7\over 2} + \gamma + {\pi^2\over3} \right)
		\hat m_D^4
\nonumber
\\
&&  \hspace{2mm}
+ {N_c \alpha_s \over 3 \pi}
\Bigg[ - {15 \over 4} + 45 \hat m_D
 - {165 \over 4}
\left( \log{\hat \mu \over 2 }  \right.
\nonumber \\ &&  \hspace{2cm} \left.
	- {36 \over 11} \log \hat m_D - 2.001 \right) \hat m_D^2
\nonumber
\\
&& \hspace{1.3cm}
+ {495\over 2}
\left( \log{\hat \mu \over 2} + {5\over22} + \gamma \right) \hat m_D^3 \Bigg] \;.
\label{Omega-NLO}
\end{eqnarray}

\vspace{2mm}
\noindent
{\em Next-to-next-to-leading~order}~:
The renormalized NNLO thermodynamic potential is
\begin{eqnarray}
\frac{\Omega_{\rm NNLO}}{{\cal F}_{\rm ideal}} &\!=\!&
	1 - {15\over4} \hat m_D^3 + {N_c\alpha_s\over3\pi} \left[ -{15\over4}  + {45\over2} \hat m_D \right.
\nonumber \\  && \hspace{4mm}	 \left.
	- {135\over2} \hat m_D^2 
	-{495\over4} \left( \log{\hat\mu\over2} + {5\over22} + \gamma\right)\!\hat m_D^3 \right]
\nonumber \\  && \hspace{2mm}
+ \left({N_c\alpha_s\over3\pi}\right)^2 \Bigg[{45\over4}{1\over\hat m_D} 
 - {165\over8} \left( \log{\hat\mu \over 2}  \right.
 \nonumber \\  && \hspace{4mm} \left.
 - {72\over11}\log{\hat m_D} - {84\over55}  
- {6\over11}\gamma  - {74\over11}{\zeta^{\prime}(-1)\over\zeta(-1)} \right.
 \nonumber \\  && \hspace{4mm} \left.
 + {19\over11}{\zeta^{\prime}(-3)\over\zeta(-3)}
\right) + {1485\over4}\left(\log{\hat\mu \over 2} - {79\over44} 
\right.
 \nonumber \\  && \hspace{4mm} \left.
 + \gamma + \log2 - {\pi^2\over11}\right)\!\hat m_D\Bigg]
\;,
\label{Omega-NNLO}
\end{eqnarray}
where $\zeta$ is the Riemann $\zeta$ function.
Note that if the leading order Debye mass, $\hat m_{D,\rm LO}^2 = N_c \alpha_s/(3\pi)$, 
is used for the Debye mass in (\ref{Omega-NNLO}) we reproduce the known expansion of the Yang-Mills
free energy up to order $\alpha_s^{5/2}$.

\vspace{2mm}
\noindent
{\em Mass Prescription}~:
The mass parameter $m_D$ in HTLpt is, 
in principle, completely arbitrary. To complete a 
calculation, it is necessary to specify $m_D$ as 
function of $\alpha_s$ and $T$.  Unfortunately, similar to NNLO HTLpt QED \cite{qednnlo} the variational
mass prescription gives a complex Debye mass.  Here we equate the Debye mass used in HTLpt with the 
hard contribution to the Debye mass obtained using dimensional reduction \cite{Braaten:1995jr},
i.e. $m_D = m_E$ giving
\begin{equation}
\frac{\hat m_D^2}{\hat m_{D,\rm LO}^2} =  1 + \frac{N_c \alpha_s }{3 \pi}
\left(\frac{5}{4} + \frac{11}{2} \gamma +\frac{11}{2} \log \frac{\hat\mu}{2} \right) \, .
\label{bnmass}
\end{equation}

\vspace{3mm}
\noindent
\textsc{Results}~:
In Figs.~\ref{fig:pressure}-\ref{fig:entropy} we show the $N_c=3$ 
pressure, entropy, and energy density scaled by their respective ideal gas limits as a 
function of $T/T_c$.  The results at LO, NLO, and NNLO
use Eq.~(\ref{bnmass}) for the Debye mass in Eqs.~(\ref{Omega-LO}),  (\ref{Omega-NLO}),
and  (\ref{Omega-NNLO}), respectively.  For the running coupling we used the three-loop 
running \cite{pdg} and varied the renormalization scale by two around $\mu = 2 \pi T$.

For the pressure, energy density, and entropy
the convergence of the successive approximations to the Yang-Mills thermodynamic
functions is improved over naive perturbation theory.  For example, using the naive
perturbative approach and comparing the full variation in both successive truncations 
and renormalization scale
variation, one finds that at $T=3\,T_c$ there is variation in the pressure 
of  $0.69 \leq {\cal P}/{\cal P}_{\rm ideal} \leq 1.32$ \cite{htlpt2},
whereas using HTLpt there is only a variation of 
$0.74 \leq {\cal P}/{\cal P}_{\rm ideal} \leq 0.95$.  Additionally, at NNLO we see that
the $\mu=2\pi T$ result for the pressure in Fig.~\ref{fig:pressure} coincides with the lattice
data down to $T \sim 3\,T_c$ and the energy density and entropy are compatible with
lattice data down to $T \sim 2\,T_c$.  

However, for all thermodynamic functions we find that the NNLO results represent a significant
correction to the LO and NLO curves.  This is unexpected since the LO and NLO bands  
overlap with one another at all temperatures shown.  In addition, in NNLO HTLpt QED 
\cite{qednnlo} such a large correction was not observed.
For SU(3) Yang-Mills, in order for the LO, NLO, and NNLO bands to overlap one 
must go to temperatures $T \gsim 7\,T_c$.  One may wonder if there is 
an error in the NNLO thermodynamic potential; however, we are confident in this
result because firstly, we reproduce the known perturbative expansion to order $\alpha_s^{5/2}$ 
in the weak-coupling limit using Eq.~(\ref{Omega-NNLO})
and secondly, there were highly non-trivial cancellations of 
divergences using only systematically predicted counterterms during the renormalization procedure.
We note that in scalar theories the corresponding reorganization has been
pushed to N$^3$LO \cite{spt2} where it has been shown that the N$^3$LO result
is between the NLO and NNLO results, indicating an excellent pattern of convergence.

\vspace{3mm}
\noindent
\textsc{Conclusions and Outlook}~:
In this Letter we have presented a new result for the NNLO thermodynamic
functions for SU($N_c$) Yang-Mills theory using the HTLpt reorganization.  We 
compared our predictions 
with lattice data for $N_c=3$ and found that HTLpt is consistent with available lattice
data down to approximately $T \sim 3\,T_c$ in the case of the pressure and $T \sim 2\,T_c$ in the case 
of the energy density and entropy.  These results are in line with expectations since
below $T \sim 2-3\,T_c$ a simple ``electric'' quasiparticle 
approximation breaks down due to magnetic/nonperturbative effects.  

We found that at NNLO the variational solution for
the Debye mass becomes complex and, as a result,
we chose instead a NLO perturbative mass prescription.  The complexity
of the variational solution may be due to the truncation in $m_D/T$; however,
checking this hypothesis will require future work.  We also found
that there was a large correction going from NLO to NNLO indicating that
perhaps the result is not fully converged.  Unfortunately, it is impossible
to say how much a N$^3$LO calculation will affect things, so
again future work is required.

\vspace{3mm}
\noindent
\textsc{Acknowledgments}~:
N. S. was supported by the Frankfurt International Graduate School for Science. 
M. S. was supported in part 
by the Helmholtz International Center for FAIR Landesoffensive zur Entwicklung 
Wissenschaftlich-\"Okonomischer Exzellenz program.

\end{document}